\documentclass[preprint,showpacs,prb,floatfix,superscriptaddress,
citeautoscript,cite]{revtex4}
\usepackage{graphicx}
\usepackage{color}

\begin{document}

\title{Vacancy Formation and Oxidation Characteristics of Single Layer 
TiS$_{3}$}

\author{F. Iyikanat}
\email{fadiliyikanat@iyte.edu.tr}
\affiliation{Department of Physics, Izmir Institute of Technology,
35430 Izmir, Turkey}

\author{H. Sahin}
\email{hasan.sahin@uantwerpen.be}
\affiliation{Department of Physics, University of Antwerp, Groenenborgerlaan 171, B-2020 
Antwerpen, Belgium}

\author{R. T. Senger}
\affiliation{Department of Physics, Izmir Institute of Technology,
35430 Izmir, Turkey}

\author{F. M. Peeters}
\affiliation{Department of Physics, University of Antwerp, Groenenborgerlaan 171, B-2020 
Antwerpen, Belgium}

\date{\today}

\pacs{73.20.Hb, 82.45.Mp, 73.61.-r, 73.90.+f, 68.55.Ln}

\begin{abstract}

The structural, electronic and magnetic properties of pristine, defective and 
oxidized monolayer TiS$_{3}$ are investigated using first-principles calculations 
in the framework of density functional theory. We found that a single layer of TiS$_{3}$ 
is a direct band gap semiconductor and the bonding nature of the crystal is fundamentally 
different from other transition metal chalcogenides. The negatively charged surfaces of single 
layer TiS$_{3}$ makes this crystal a promising material for lubrication applications. The 
formation energies of possible vacancies, i.e. S, Ti, TiS and double S, are investigated via 
total energy optimization calculations. We found that the formation of a single S vacancy was the 
most likely one among the considered vacancy types. While a single S vacancy results in a 
nonmagnetic, semiconducting character with an enhanced band gap, other vacancy types induce 
metallic behavior with spin polarization of $0.3-0.8$ $\mu_{B}$. The reactivity of pristine 
and defective TiS$_{3}$ crystals against oxidation was investigated using conjugate gradient 
calculations where we considered the interaction with atomic O, O$_{2}$ and O$_{3}$. While 
O$_{2}$ has the lowest binding energy with $0.05-0.07$ eV, O$_{3}$ forms strong 
bonds stable even at moderate temperatures. The strong interaction ($3.9-4.0$ eV) between 
atomic O and TiS$_{3}$ results in dissociative adsorption of some O-containing molecules. 
In addition, the presence of S-vacancies enhances the reactivity of the surface with atomic 
O whereas, it had a negative effect on the reactivity with O$_{2}$ and O$_{3}$ molecules.

\end{abstract}

\maketitle

\section{Introduction}

The wide range of different structural forms and the tunability of their electronic  
properties, has made two-dimensional ultra-thin materials popular in the field of 
condensed matter physics. In the last decade, following the synthesis of 
graphene,\cite{Novoselov, Geim} layered crystals of transition metal dichalcogenides 
(TMDs) have attracted tremendous interest owing to their remarkable electronic, 
mechanical and optical properties.\cite{Wang, Xu, Huang} Depending on the 
combination of metal and chalcogen atoms, TMDs may display metallic, 
semimetallic, semiconducting and even superconducting behavior. \cite{Yoffe, 
Zhang, Ang, Splendiani, Ataca, Salvo} Various experimental methods to thin down 
TMDs to monolayers\cite{Wang1, Nicolosi, Chhowalla} and a number 
of theoretical studies revealing the possibility of tuning their properties 
have already been reported.\cite{Mak, Iyikanat, Horzum, Zeng} In addition to 
TMDs (MoX$_{2}$ X= S, Se, Te) having graphene-like crystal structure, recent efforts 
have also led to the synthesis of novel TMDs with entirely different crystal 
structures. Titanium trisulphide (TiS$_{3}$) is one of the most recent examples of such 
layered transition metal chalcogenides. 

The atomic structure belongs to the space group P2$_{1}$/m and TiS$_{3}$ 
crystallizes in bundles of molecular chains that are formed by trigonal prisms, where the metal atoms 
occupy the centers of the prisms.\cite{Brattas, Furuseth} Finkman \textit{et 
al.}\cite{Finkman} have shown that TiS$_{3}$ crystals are semiconducting with extrinsic 
n-type conductivity that have room temperature mobility of $30$ cm$^{2}/$(V 
sec). Recently, strong nonlinearity of the current$-$voltage characteristics has
been reported by Gorlava \textit{et al.}\cite{Gorlova, Gorlova1} In contrast to many other 
layered TMDs which exhibit an indirect to direct band gap transition at the monolayer limit, 
TiS$_{3}$ exhibits a direct band gap even for a width of hundreds of layers. Direct optical transitions with a band 
gap of $1.10$ eV have been detected in thin films of TiS$_{3}$.\cite{Ferrer} 
Moreover, these thin films show photocurrent response to white light 
illumination.\cite{Ferrer1} 

Using high-resolution TEM for defective crystals, a metal-to-insulator 
transition with charge localization below \textit{T}$_{MI}$ $\approx$ $325$ K has 
been proposed by Guilmeau \textit{et al.}\cite{Guilmeau} Their studies revealed 
that TiS$_{3}$ has low thermal conductivity and a large absolute value of the 
Seebeck coefficient at high temperatures. In 
a recent study by Island \textit{et al.}\cite{island} field effect transistors 
(NR-FET) have been fabricated at room 
temperature by isolating few-layer TiS$_{3}$ nanoribbons. The electron mobility of few-layer 
TiS$_{3}$ was found to be $2.6$ cm$^{2}/$(V sec), and exhibits n-type semiconductor behavior 
with ultrahigh photoresponse and fast switching times. Furthermore, Island \textit{et al.} \cite{island-1}
have isolated single-layer TiS$_{3}$ by mechanical exfoliation.

In this study, motivated by the latest experimental\cite{island, island-1, Barawi, Gorlova2, Gorlova3, Tanibata} 
and theoretical\cite{Wu,Jin} studies on TiS$_{3}$, we investigated: (i) structural and electronic properties of 
single layer TiS$_{3}$, (ii) formation of vacancies in the pristine material and 
their influence on electronic properties, and (iii) environmental stability of its 
surface against oxidation. The paper is organized as follows. In Sec. \ref{sec2}, we present the details 
of the computational methodology. Geometric, electronic and magnetic 
properties of defect-free TiS$_{3}$ and of various vacancy defects 
are given in Sec. \ref{sec3} and Sec. \ref{sec4}, respectively. The response 
of pristine and defective TiS$_{3}$ to oxidation is analyzed in Sec. 
\ref{sec5}. Results are discussed in Sec.\ref{sec6}.

\section{Computational Methodology}\label{sec2}
All the calculations were performed within the spin-polarized density functional 
theory (DFT) using projector-augmented-wave potentials (PAW) and a plane-wave 
basis set as implemented in the Vienna ab initio simulation package (VASP).\cite{Kresse, Kresse1} 
The cutoff energy for the plane-waves was chosen to be $500$ eV. 
Perdew-Burke-Ernzerhof’s (PBE) version of generalized gradient approximation (GGA) \cite{Perdew} 
was used for the description of the exchange correlation functional. For a better 
approximation of bandgap values, 
underestimated by PBE functional, the HSE06 functional was also used.\cite{Heyd} In 
the HSE06 approach, the fraction of the Hartree-Fock exchange and the screening 
parameter were set to $\alpha$ $=$ $0.25$ and $0.2$ \AA\/$^{-1}$, respectively. 

Spin-polarized calculations were performed in all cases and atomic charges were 
calculated using the Bader charge population analysis.\cite{Henkelman} 
The Gaussian smearing method was employed for total energy 
calculations with a width of $0.01$ eV. Density functional theory plus the long-range dispersion correction 
(DFT+D2) method was used to calculate the non-local correlation 
energies.\cite{Grimme} C$_{6}$ values of Ti and S atoms were $10.800$ and $5.570$, respectively. 
$1.562$ and $1.683$ were used as the vdW radius of Ti and S atoms, 
respectively.\cite{Grimme}

Geometric structures of the perfect, vacancy defected and oxidized monolayer TiS$_{3}$ 
were fully optimized to minimize each component of the inter-atomic Hellmann-Feynman 
forces until a precision of $10^{-4}$ eV/\AA\ was reached. The pressure in the 
unit cell was kept below $1$ kBar. The convergence criterion for energy was 
chosen to be $10^{-5}$ eV between two consecutive steps. The conjugate gradient 
method was used to compute lattice constants and total energies. In all cases, 
lattice parameters were optimized along the \textit{a$_{1}$} 
and \textit{a$_{2}$} directions. In order to hinder interlayer interaction 
within the periodic images, a vacuum spacing of $16$ \AA\ 
between adjacent layers was chosen. The lateral distance between vacancies 
(adsorbed atoms) was at least $10$ \AA\ in order to eliminate the interaction 
between vacancies (adsorbed atoms) in neighboring supercells. For the 
unitcell of TiS$_{3}$, $19\times19\times1$ k-point mesh was used in PBE 
calculations, whereas due to high computational cost, the k-point mesh was reduced to 
$5\times5\times1$ in HSE calculations. For the supercell calculations, the same k-point 
mesh of $3\times3\times1$ was used for both PBE and HSE. The vacancies were obtained by 
removing the considered atoms from a $3\times3$ supercell which consist of $18$ Ti atoms 
and $54$ S atoms. 

The calculated formation energies of the vacancies was obtained 
from $E_{F} = E_{M-A} - E_{M} + nE_{A}$, where $E_{F}$ is the formation 
energy of the relevant vacancy, $E_{M-A}$ is the total energy of the supercell with 
vacancy, $n$ is the number of removed atoms, $E_{A}$ is the energy of 
the removed atom and $E_{M}$ is the energy of monolayer TiS$_{3}$. Binding energies 
of O, O$_{2}$ and O$_{3}$ were calculated for the most favorable adsorption 
sites. These binding energies were calculated from the expression $E_{B} = 
E_{M+A}(E_{D+A}) - E_{M}(E_{D}) - mE_{A}$ , where E$_{B}$ is the binding energy 
of the adsorbed atom on pristine (S defected) TiS$_{3}$, $m$ is the number 
of adsorbed atoms, E$_{M}$ (E$_{D}$) and E$_{M+A}$ (E$_{D+A}$) are the total 
energy of pristine (S defected) and the adsorbed atom-pristine (adsorbed 
atom-S-defected) system, respectively.

\begin{figure}
\includegraphics[width=8.5cm]{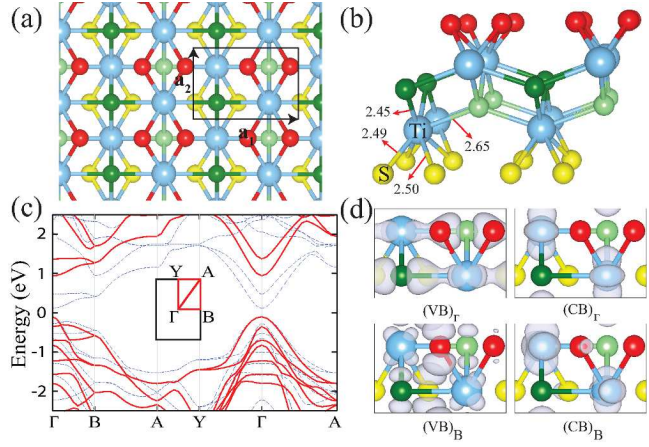}
\caption{\label{structures-1}
(Color online) (a) Top view of monolayer TiS$_{3}$ and its unit cell shown by a 
rectangle. For a better top view, S atoms in different layers are 
presented by different colors. (b) Tilted side view of 2x2x1 supercell of 
monolayer TiS$_{3}$. (c) Blue dashed lines and red lines illustrate PBE and 
HSE06 results for the electronic band diagram of monolayer TiS$_{3}$, respectively. 
(d) Band decomposed charge densities of valance band (VB) and conductance 
band (CB) at $\Gamma$ and B.}
\end{figure}

\section{Pristine Single Layer T\lowercase{i}S$_{3}$}\label{sec3}

The unit cell of monolayer TiS$_{3}$ is a rectangular prism and is composed of $2$ 
Ti atoms and $6$ S atoms. The coordination of these atoms are illustrated 
by the top and tilted-side views in Figs. \ref{structures-1}(a) and \ref{structures-1}(b), 
respectively. The monolayer TiS$_{3}$ is composed of chain-like structures consisting of 
trigonal prisms with the metal atom occupying 
the centers of these prisms. In this phase, chains are parallel to the 
\textit{a$_{2}$} and they form layers in the \textit{a$_{1}$}\textit{a$_{2}$} 
plane, which are coupled with each other by the van der Waals interaction to 
form the bulk layered structure of TiS$_{3}$. Our calculations show 
that monolayer TiS$_{3}$ has lattice vectors \textit{a$_{1}$} $=$ $4.99$ \AA\ 
and \textit{a$_{2}$} $=$ $3.39$ \AA\/. These lattice parameters are comparable to the 
experimental bulk results \textit{a$_{1}$} $=$ $4.958$ \AA\/, 
\textit{a$_{2}$} $=$ $3.4006$ \AA\/, \textit{a$_{3}$} $=$ $8.778$ \AA\/, and $\beta = 97.32$.\cite{Brattas, Furuseth} 
As shown in Fig. \ref{structures-1}(b) the bond distances of Ti atom and two different surface 
S atoms (red or yellow S atoms) are $2.49$ \AA\ and $2.50$ \AA\/, respectively. At the same time, 
these two surface S atoms are two of the three base S atoms of the trigonal prisms. 
The bond distance of Ti atom and third base S atom (dark or light green S atoms) of the 
trigonal prism is $2.45$ \AA\/. The atom-atom distance of Ti atom and S atom (light or 
dark green S atoms) of neighboring prism is $2.65$ \AA\/.

\begin{figure}
\includegraphics[width=8.5cm]{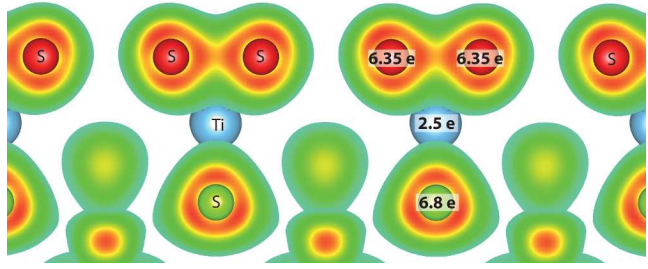}
\caption{\label{structures-2}
(Color online) Cross-sectional plane view of the charge densities, with two surface, and one inner 
S atom placed at this plane. Valance charges on inner and outer S atoms are also shown. 
Color code of atoms is the same as in Fig. \ref{structures-1}. Electron density increases from green to red. }
\end{figure}

Charge density analysis is an efficient way to discuss the character of the interatomic interactions and bonding. 
In Fig. \ref{structures-2}, we present a cross section of the 3D total charge density. Bader charge 
analysis shows that in parallel to their electronegativity values a significant amount of charge is transferred from Ti 
to the S atoms. While two S atoms at the surface (shown by red balls) share $0.7$ electrons donated by the underlying Ti 
atom, $0.8$ electron transfer occurs from Ti to S  atom in the middle of the crystal (shown by light green balls). 
Thus, Ti-S bonds between surface and inner S atoms have an entirely different character. As is also delineated in 
Fig. \ref{structures-2} the bond between Ti and inner S atoms is constructed through $0.8$ electron transfer and 
therefore it has an ionic character. However, the bonds between Ti and surface S atoms are constructed through relatively 
less electron transfer and hence it has mostly covalent character.  It is also seen from Fig. \ref{structures-2} that each 
surface atom interacts with neighboring S atoms while there is no contact between the inner S atoms. Therefore one expects 
anisotropic electronic and transport properties due to the varying character of the surface states along \textit{a$_{1}$} 
and \textit{a$_{2}$} directions. It is also worth mentioning that the negatively charged surface of monolayer TiS$_{3}$ may 
find interesting applications  such as nanoscale lubricants and charged coatings.

Calculated band diagrams of pristine TiS$_{3}$ using PBE and HSE06 
methods are shown in Fig. \ref{structures-1}(c). It is known that often the 
values, obtained with the PBE functional underestimates the energy band gaps of semiconductors. 
The band gap of TiS$_{3}$ is calculated to be $0.23$ eV by 
using the PBE approximation. However, including the HSE06 correction one 
gets a $1.05$ eV direct band gap at the $\Gamma$ point, which is in good 
agreement with the experimental value of $1.10$ eV, for 
few-layer TiS$_{3}$.\cite{Ferrer, Ferrer1} For further analysis of the 
band structure of TiS$_{3}$, charge densities of the valance band maximum (VB) 
and the conduction band minimum (CB) at the $\Gamma$ and the B high-symmetry 
points are shown in Fig. \ref{structures-1}(d). It is seen that the VB is composed 
of $2p_{x}$ orbitals of S and $3d_{xz}$ orbitals of Ti and they form a strong 
bond, whereas the main 
contribution to the CB edge comes from $3d_{y^{2}-z^{2}}$ orbitals 
of Ti atom at the $\Gamma$ point. At the B point, $p_{y}$ orbitals of S atoms 
dominant while $3d_{xy}$ orbitals of Ti atom contribute slightly to the VB. In particular, 
$p$ orbitals of surface S atoms are larger than $p$ orbitals of the inner S atoms. Thus, the main 
contribution to the VB is dominated by $p_{y}$ orbitals of the surface S atoms. 
Like at the $\Gamma$ point, while the S atoms do not contribute significantly, 
all the contribution comes from the $3d_{y^{2}-z^{2}}$ orbitals of the Ti atom to 
the CB. In the trigonal prisms every Ti atom shares $4$ valence electrons with $3$ 
base S atoms and $1$ neighboring chain S atom. Due to the lack of any unsaturated 
orbitals, monolayer TiS$_{3}$ has a nonmagnetic ground state.

\section{Defective Single Layer T\lowercase{i}S$_{3}$}\label{sec4}

\subsection{S-vacancy}

\begin{figure}
\includegraphics[width=8.5cm]{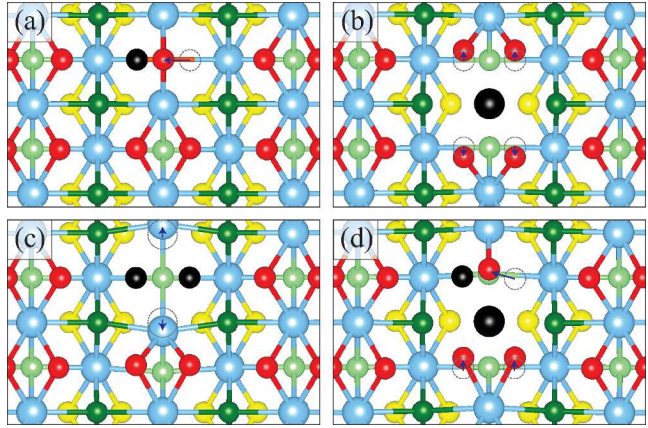}
\caption{\label{structures-3}
(Color online) Top view of relaxed monolayer TiS$_{3}$ with (a) S-vacancy, (b) 
Ti-vacancy, (c) Double S-vacancy, (d) TiS-vacancy. Black atoms illustrate removed 
atoms, dashed circles show initial position of the displaced atom. Color code of 
the other atoms is the same as in Fig. \ref{structures-1}. The direction of displaced atoms 
are depicted by blue arrows.}
\end{figure}

\begin{table}
\caption{\label{table-1} Lattice parameters, formation energies, magnetic moments, electronic 
characteristics and band gaps of $3\times3$ supercell of monolayer TiS$_{3}$ and its few defected forms 
calculated using PBE method. HSE06 results for the band gap of semiconductors are also given.}
\begin{tabular}{lcccccccccc}
\hline\hline
         & \textit{a$_{1}$} / \textit{a$_{2}$} & E$_{F}$ & m & Electronic & Band Gap \\
         & ( \AA\ )& (eV) & ($\mu$B) & Characteristic & (eV) \\
\hline
Pristine TiS$_{3}$   & $14.98$ / $10.18$ & - & $0.0$ & Semiconductor & 0.23(PBE)/1.05(HSE06) \\
S Vacancy  & $14.90$ / $10.17$ & $3.58$ & $0.0$ & Semiconductor & 0.28(PBE)/0.89(HSE06) \\
Ti Vacancy  & $15.05$ / $10.14$ & $12.00$ & $0.5$ & Metal & - \\
Double S Vacancy  & $14.96$ / $10.18$ & $8.48$ & $0.8$ & Metal & - \\
Ti, S Vacancy  & $14.90$ / $10.13$ & $16.15$ & $0.3$ & Metal & - \\
\hline\hline
\end{tabular}
\end{table}
The first defective structure we consider is a single S vacancy in TiS$_{3}$ monolayer. 
To compare the geometric structure of monolayer TiS$_{3}$ in the presence and absence of a vacancy, 
a $3\times3$ supercell is considered and the lattice vectors for the defect-free TiS$_{3}$ computational 
supercell are found as \textit{a$_{1}$} $= 14.98$ \AA\ and \textit{a$_{2}$} $= 10.18$ \AA\/. 
Fully relaxed geometric structure when a single S atom is removed from the surface of monolayer TiS$_{3}$ 
is shown in Fig. \ref{structures-3}(a). When S vacancy is introduced the lattice vectors of $3\times3$ 
supercell change and become \textit{a$_{1}$} $= 14.90$ \AA\ and \textit{a$_{2}$} $= 10.17$ \AA\/. Thus, 
the presence of the S vacancy leads to a minute shrinkage of the lattice vectors of TiS$_{3}$. As can be seen 
from Fig. \ref{structures-3}(a), the nearest surface S atom to the vacancy follows the direction of the arrow, 
localizes on top of the inner S atom (light green) and forms reconstructed bonds with the nearest Ti atoms with 
a bond length of $2.32$ \AA\/. Bader charge analysis tells us that, the total charge of this atom is increased 
by $0.4$ electrons when removing the nearest neighbor S atom. The charges of other atoms do not change significantly 
with the removal of the surface S atom. Our calculated results show that the formation energy of an S-vacancy is 
$3.58$ eV.

To calculate the electronic properties of S-vacancy defected monolayer TiS$_{3}$ in addition to PBE, 
HSE06 method is also used and the results are listed in Table \ref{table-1}. Moreover, the density of states (DOS) of 
pristine and S-atom-removed TiS$_{3}$ calculated by PBE method are shown in Figs. \ref{structures-4}(a) 
and \ref{structures-4}(b), respectively. As can be seen from the figures and Table \ref{table-1}, removing one S atom 
from the surface of TiS$_{3}$ does not make any notable effect on the electronic structure of TiS$_{3}$. The monolayer 
conserves its semiconductor character. Due to the reconstruction of the S atom and its binding with 2 Ti atoms, there 
are no unsaturated bonds. Thus, the nonmagnetic character of the TiS$_{3}$ is preserved during the formation of an S-vacancy.

\subsection{Ti-vacancy}

Relaxed geometric structure when a single Ti atom is extracted from TiS$_{3}$ is 
illustrated in Fig. \ref{structures-3}(b). A significant surface reconstruction is 
observed after geometric relaxation. This figure shows that, when one Ti atom is taken 
out, the displacement of the inner S atoms that were attached to it, is not significant,
nevertheless they loose 0.1 electrons. However, the four surface S atoms are released, 
they move towards the other Ti atoms and they lose approximately 0.2 electrons. This gives 
rise to an expansion in the lattice parameter \textit{a$_{1}$}, which after relaxation becomes 
\textit{a$_{1}$} $= 15.05$ \AA\ and  \textit{a$_{2}$} $= 10.14$ \AA\/. 
Hence, removal of Ti atom leads to an increase in \textit{a$_{1}$}, 
whereas a decrease in \textit{a$_{2}$}. The formation energy of Ti-vacancy is $12.00$ eV. 
Defects with high formation energies are unlikely to form, and therefore an S-vacancy
will have a much higher probability to form as compared to a Ti-vacancy.

Unlike S vacancy, Ti vacancy has a major effect on the electronic structure of 
TiS$_{3}$, it loses its semiconductor character and becomes metallic. 
As shown in Fig \ref{structures-4}(c) the states orginating from Ti vacancy are near the 
VB with three peaks of gap states arising from the mixture of the orbitals of the neighboring S 
and Ti atoms. This leads the possibility of p-type doping if Ti vacancy can be created. As 
seen from Fig. \ref{structures-4}(c), at the Fermi level, there is a slight spin polarization and 
the contribution from the S atom is more than that of the Ti atom. As given in Table \ref{table-1}, 
TiS$_{3}$ monolayer with a Ti-vacancy exhibits a magnetic ground state and the 
calculated magnetic moment is $0.5$ $\mu_{B}$$/$supercell.

\subsection{Double S-vacancy}

Possible two-atom-vacancy structures of TiS$_{3}$ are also investigated. First, the structure of 
double S atoms removed from the surface of TiS$_{3}$ is studied. Fig. \ref{structures-3}(c) presents its 
fully relaxed configuration. Compared to the one S vacancy situation, the lattice is more deformed. 
The double vacancy of surface S atoms leads to a reconstruction of the bonds between the S atoms 
of neighboring prisms (dark green one) and the Ti atoms closest to the vacancies. These Ti atoms 
become more strongly bonded to the surface S atoms. However, compared to the defect-free structure, when 
two S atoms are removed the lattice parameters are slightly changed. Bader charge analysis shows that, 
removing two surface S atoms does not affect the charges of the remaining Ti atoms. However, $0.7$ 
excess electrons of the removed two S atoms are shared by neighbor S atoms. The formation energy of double 
S-vacancy is $8.48$ eV. Thus, removing two S atoms from the TiS$_{3}$ surface is more probable than 
removing one Ti atom.  

\begin{figure}
\includegraphics[width=8.5cm]{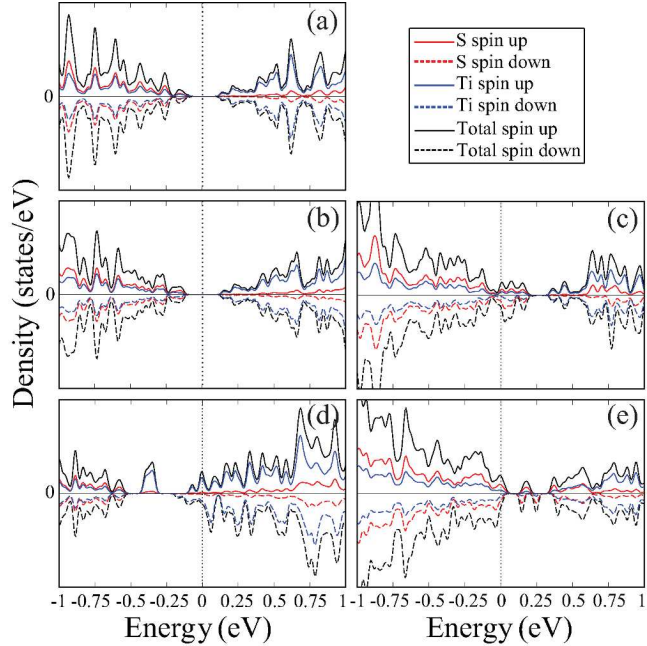}
\caption{\label{structures-4}
Density of states of (a) pristine TiS$_{3}$, with (b) S-vacancy, (c) Ti-vacancy, (d) Double S-vacancy, and 
(e) TiS-vacancy.  
}
\end{figure}

The calculated density of states of monolayer TiS$_{3}$ with double S-vacancy is presented in 
Fig. \ref{structures-4}(d). When the second S atom is taken out from the surface, TiS$_{3}$ exhibits metallic 
character. As seen from Fig. \ref{structures-4}(d) the main contribution comes from the Ti atoms and TiS$_{3}$ 
with double S-vacancy has an asymmetric DOS at the Fermi level. This asymmetric DOS at the 
Fermi level leads to a magnetic ground state with the magnetic moment value of $0.8$ $\mu_{B}$$/$supercell.

\subsection{TiS-vacancy}
Lastly, we considered the defected structure of TiS$_{3}$ with a TiS-vacancy. Fig. \ref{structures-3}(d) 
shows that in the presence of TiS-vacancy, the position of the closest surface S 
atom to the removed S atom is significantly changed. This atom follows the direction of the 
arrow and becomes located at the top of the inner S atom (light green) with charge 6.6 electrons. 
Other two surface S atoms move toward the vacancies and their charges reduce to 6.0 and 6.2 electrons. The 
lattice parameters reduce to the values \textit{a$_{1}$} $= 14.90$ \AA\ and \textit{a$_{2}$} $= 10.13$ \AA\/. 
Compared to the other vacancy types its formation energy value is the highest with the value $16.15$ eV.

DOS diagram of TiS$_{3}$ with TiS-vacancy is illustrated in Fig. \ref{structures-4}(e). Like Ti and 
double S-vacancies, this vacancy type also leads to a metallic character. But, 
unlike the double S-vacancy, Fermi level consists of orbitals of S atom. As given 
in Table \ref{table-1}, the magnetic moment of this case is $0.3$ $\mu_{B}$$/$supercell.

\section{Oxidation of Pristine and Defective T\lowercase{i}S$_{3}$}\label{sec5}

It is well-known that two-dimensional ultra-thin structures such as 
graphene, MoS$_{2}$, phosphorene etc. are prone to oxidation.\cite{Vinogradov, Wang2, Yamamoto} 
Therefore, the search for structural and environmental stability of TiS$_{3}$ is of 
importance. After the investigation of possible defect types in monolayer 
TiS$_{3}$, in this section we address the oxidation process and the role 
of vacancies in that process.

\begin{figure}
\includegraphics[width=8.5cm]{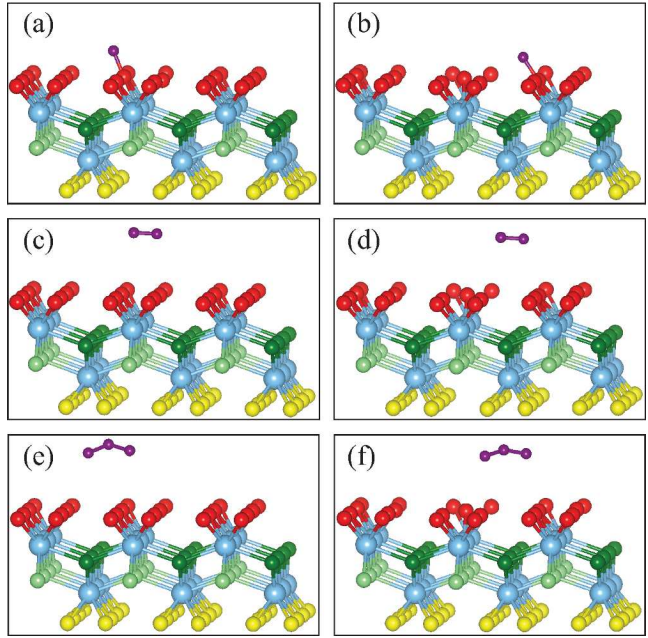}
\caption{\label{structures-5}
(Color online) Tilted side view of (a) O, (c) O$_{2}$ and (e) O$_{3}$ adsorbed on pristine
TiS$_{3}$. (b) O, (d) O$_{2}$ and (f) O$_{3}$ adsorbed on an S-vacant TiS$_{3}$.
}
\end{figure}

\subsection{O atom adsorption on pristine and defective TiS$_{3}$}

During experiments in highly oxidative conditions, the presence of Oxygen atom is 
inevitable and atomic O can be considered as a powerful tool to functionalize the surface 
of TiS$_{3}$. Thus, we start to investigate the oxidation with atomic O on pristine 
TiS$_{3}$. As shown in Fig. \ref{structures-5}(a), the optimized structures show that, 
an O atom is adsorbed by an S atom on the pristine TiS$_{3}$ surface. 
The distance between O and S atoms is about $1.49$ \AA. The presence of the O atom does not 
have any significant effect on the lattice structure of TiS$_{3}$. O atom binds strongly to 
the S atom whose bond has an ionic character. The amount of charge transfer between O and 
S atoms is about $1.3$ electrons from S to the O atom. The binding energy of the O atom on TiS$_{3}$ 
crystal is $-3.89$ eV. When O atom binds to the TiS$_{3}$ surface, it loses its magnetic moment 
and the whole system does not exhibit any net magnetic moment.

The vacancy calculations have shown that the most probable defect type of TiS$_{3}$ is the 
S atom vacancy. Since the O atom readily binds to the pristine TiS$_{3}$ surface, the 
effect of the S-vacancy on the adsorption of atomic O is in order. 
Fig. \ref{structures-5}(b) shows that, instead of being the nearest S atom to the 
vacancy, O atom prefers to bind to the other surface-S atom from tilted-top 
site. Compared to the pristine case, the presence of the S vacancy increases the 
binding energy of O atom from $-3.89$ eV to $-4.01$ eV. The bond distance 
between O and S atoms is $1.48$ \AA\/. Unlike the pristine case, O atom distorts 
the lattice of defective TiS$_{3}$ and the new lattice vectors are expanded to 
\textit{a$_{1}$} $= 14.93$ 
\AA\ and \textit{a$_{2}$} $= 10.19$ \AA\/. O atom binds ionically to the surface S 
atom and $1.1$ electrons transfer from surface S atom to the O atom. Like in the 
pristine case, the net magnetic moment of the whole system is zero.

We also investigate substitutional adsorption of O atom on the surface of 
single layer TiS$_{3}$. The formation energy of the substitutional O atom at 
S site is calculated using the formula $E_{Subs} = E_{M+O}-E_{M}-E_{O}+E_{S}$, 
where $E_{M+O}$ denotes the energy of O-doped TiS$_{3}$, $E_{M}$ is the 
energy of pristine TiS$_{3}$, E$_{O}$ and E$_{S}$ are single atom energies. 
Substitution energy of O with S is found to be $-1.38$ eV. Negative value for 
O atom indicates the spontaneous formation of substitutional doping at the S 
site for O atom. It is also seen that the presence of substitutional O does not 
change the nonmagnetic structure of TiS$_{3}$.

\subsection{O$_{2}$ molecule adsorption on pristine and defective 
TiS$_{3}$} Approximately $21$\%  of the earth atmosphere is composed of O$_{2}$ 
molecules. Thus, the binding mechanism of O$_{2}$ molecule on TiS$_{3}$ surface and 
the stability of TiS$_{3}$ in the presence of this molecule are very crucial. 
The results of our calculations show that compared to atomic O, the O$_{2}$ 
molecule binds very weakly to the TiS$_{3}$ surface and its binding energy is 
$-0.07$ eV. Fig. \ref{structures-5}(c) shows that O$_{2}$ locates $3.2$ \AA\ 
above the TiS$_{3}$ surface and it prefers to bind on top of the vicinity of S 
atom. The presence of O$_{2}$ on the TiS$_{3}$ surface does not cause any 
distortion on the lattice structure. Contrary to the single O atom case, there is 
no significant charge transfer between the surface and the molecule. Total 
magnetic moment of the system is $2\mu_{B}$, which is equal to the magnetic 
moment of an isolated O$_{2}$ molecule. 

The presence of an S-vacancy does not have any significant effect on the adsorption of 
O$_{2}$ molecule on the TiS$_{3}$ crystal. Similar to the pristine case, O$_{2}$ 
molecule binds rather weakly to S-vacancy with a binding energy of $-0.05$ eV. Compared 
to the pristine case, O$_{2}$ molecule is localized more close to the TiS$_{3}$ surface 
at a z-distance of $2.89$ \AA\/. The presence of O$_{2}$ molecule does not cause any notable 
distortion of the defected TiS$_{3}$ crystal. Like in the pristine case, there is 
almost no charge transfer between TiS$_{3}$ and the O$_{2}$ molecule and the net 
magnetic moment of O$_{2}$ molecule which is $2\mu_{B}$, does not change in the presence 
of the S-vacancy.

\subsection{O$_{3}$ molecule adsorption on pristine and defective TiS$_{3}$} To 
complete the analysis of oxidation of TiS$_{3}$, the binding mechanism of the O$_{3}$ 
molecule on the TiS$_{3}$ surface is investigated. It is found that, compared to O$_{2}$ molecule, 
O$_{3}$ molecule binds more strongly to the TiS$_{3}$ surface. Binding energy of O$_{3}$ molecule 
on TiS$_{3}$ surface is $-0.21$ eV. As seen from Fig. \ref{structures-5}(e), O$_{3}$ molecule locates 
$3.15$ \AA\ above the TiS$_{3}$ surface with two edge O atoms being closer to the surface 
compared to the middle O atom. The stability of TiS$_{3}$ is not affected by O$_{3}$. 
When the O$_{3}$ molecule is placed on the TiS$_{3}$ surface, it receives an extra $0.2$ electrons 
from the TiS$_{3}$. The net magnetic moment of the O$_{3}$ molecule adsorbed pristine TiS$_{3}$ 
is zero.

Finally, the adsorption of O$_{3}$ molecule on an S-vacancy is investigated. As seen from Fig. \ref{structures-5}(f), 
O$_{3}$ molecule locates approximately $2.95$ \AA\ above the TiS$_{3}$ surface. In the S vacancy case the end 
O atoms are placed closer to the surface compared to the middle O atom. Presence of S vacancy slightly affects 
the binding energy of O$_{3}$ molecule which is equal to $-0.20$ eV. O$_{3}$ molecule does not make any significant 
effect on the lattice structure of TiS$_{3}$ with an S-vacancy. The charge transfer from the surface to the molecule 
is about $0.2$ electrons. The adsorbed O$_{3}$ molecule on an S-vacancy does not possess any net magnetic moment.

\section{Conclusion}\label{sec6}
In this paper, a detailed analyses of structural, electronic and magnetic properties 
of pristine, defective and oxidized structures of monolayer TiS$_{3}$ are presented 
by using first-principles calculations. In addition to the PBE version of GGA, the HSE06 
form of hybrid functionals were also used to describe the exchange-correlation density 
functional. Electronic structure calculations using HSE06 hybrid functional 
indicated that monolayer TiS$_{3}$ is a direct bandgap semiconductor with a band 
gap of $1.05$ eV. Our calculations also revealed interesting bonding nature of the 
monolayer TiS$_{3}$ crystal that has ionic character inside and covalent character 
for surface atoms. The negatively charged surface of the crystal may also find some 
interesting applications such as nanoscale lubricants and charged coatings.

Among various vacancy defects including S, Ti, TiS and double 
S vacancies, the single S vacancy has the lowest formation energy. While the S vacancy leads to 
an opening in the band gap, other vacancies result in metallicity in single layer 
TiS$_{3}$. Pristine and S vacancy defected TiS$_{3}$ does not posses any net magnetic 
moment, whereas other considered vacancies are magnetic. Our DFT oxidation studies 
revealed that, TiS$_{3}$ readily oxidizes with atomic O. Moreover, it is found that, 
oxidation of TiS$_{3}$ with O$_{3}$ is most likely to occur, while oxidation with O$_{2}$ 
is less favorable on pristine and S defected TiS$_{3}$ surface.
S vacancy has a slightly negative effect on the adsorption of O$_{2}$ and O$_{3}$ 
molecules on TiS$_{3}$ surface, however it has a favoring effect on the adsorption 
of atomic O.

\begin{acknowledgments} 
This work was supported by the Flemish Science Foundation (FWO-Vl)
and the Methusalem foundation of the Flemish government.
Computational resources were provided by TUBITAK ULAKBIM,
High Performance and Grid Computing Center (TR-Grid e-Infrastructure),
and HPC infrastructure of the University of Antwerp (CalcUA) a
division of the Flemish Supercomputer Center (VSC), which is funded by
the Hercules foundation. H.S. is supported by a FWO Pegasus Marie
Curie Fellowship. F.I., H.S. and R.T.S. acknowledge the support from 
TUBITAK through project 114F397.
\end{acknowledgments}

\begin{figure}[h]
\includegraphics[width=8.5cm]{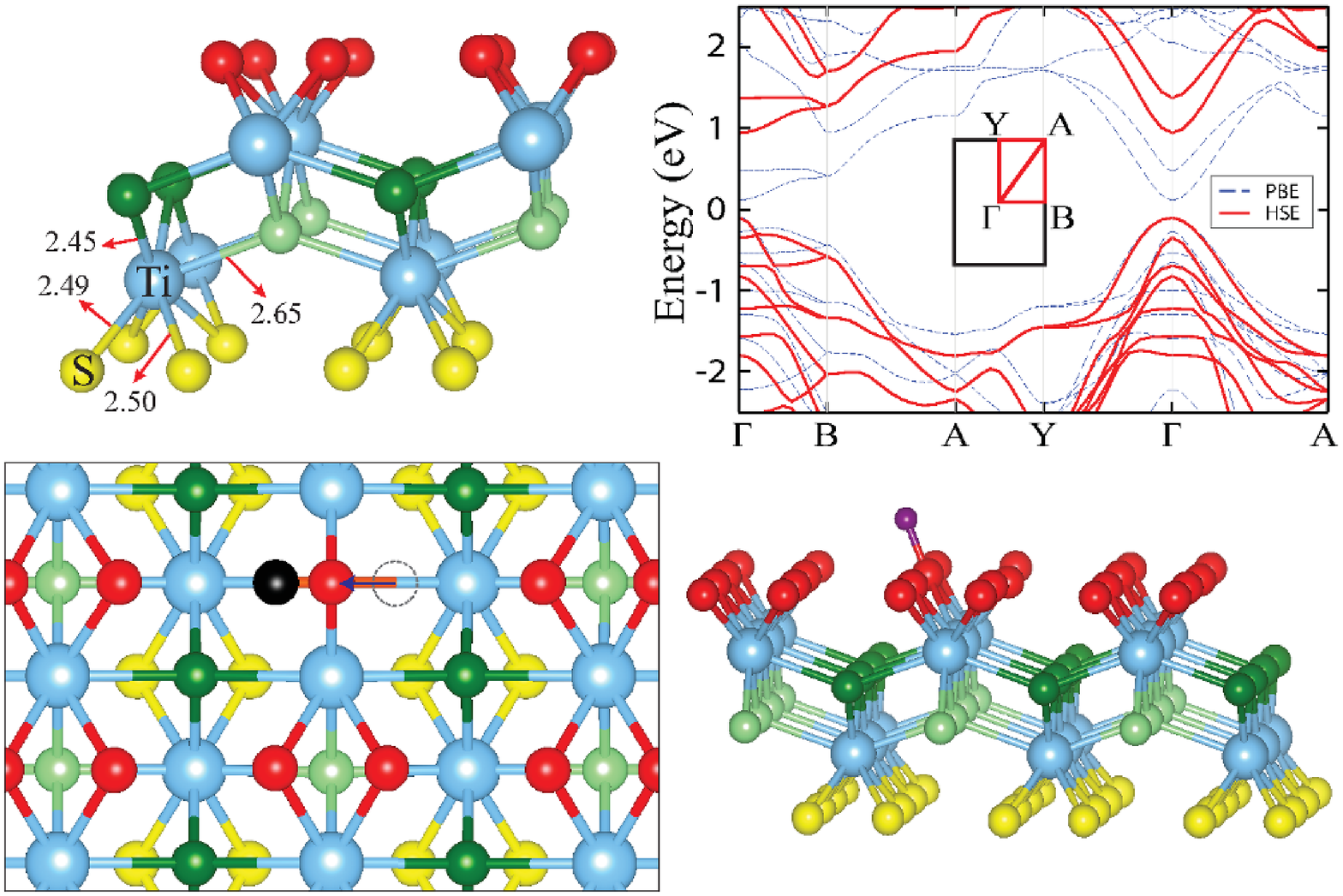}
\caption{\label{toc}
For Table of Contents Only.}
\end{figure}

\end{document}